# Estimation of Global Building Stocks by 2070: Unlocking Renovation Potential

Shufan Zhang [1], Minda Ma [2 *, 6], Nan Zhou [6 *], Jinyue Yan [3], Wei Feng [4, 5, 6], Ran Yan [1], Kairui You [7], Jingjing Zhang [6], Jing Ke [6]

1. School of Management Science and Real Estate, Chongqing University, Chongqing, 400045, PR China
2. School of Architecture and Urban Planning, Chongqing University, Chongqing, 400045, PR China
3. Department of Building Environment and Energy Engineering, The Hong Kong Polytechnic University, Kowloon, Hong Kong, PR China
4. Institute of Technology for Carbon Neutrality, Shenzhen Institute of Advanced Technology, Chinese Academy of Sciences, Shenzhen, 518055, PR China
5. Faculty of Material Science and Energy Engineering, Shenzhen Institute of Advanced Technology, Shenzhen, 518055, PR China
6. Building Technology and Urban Systems Division, Energy Technologies Area, Lawrence Berkeley National Laboratory, Berkeley, CA 94720, United States
7. School of Management and Economics, Beijing Institute of Technology, Beijing, 100081, PR China

- Corresponding author: Dr. Minda Ma, Email: maminda@lbl.gov
  Homepage: https://buildings.lbl.gov/people/minda-ma

  Corresponding author: Dr. Nan Zhou, Email: nzhou@lbl.gov
  Homepage: https://buildings.lbl.gov/people/nan-zhou

- Lead contact: Dr. Minda Ma (maminda@lbl.gov)

# GRAPHICAL ABSTRACT

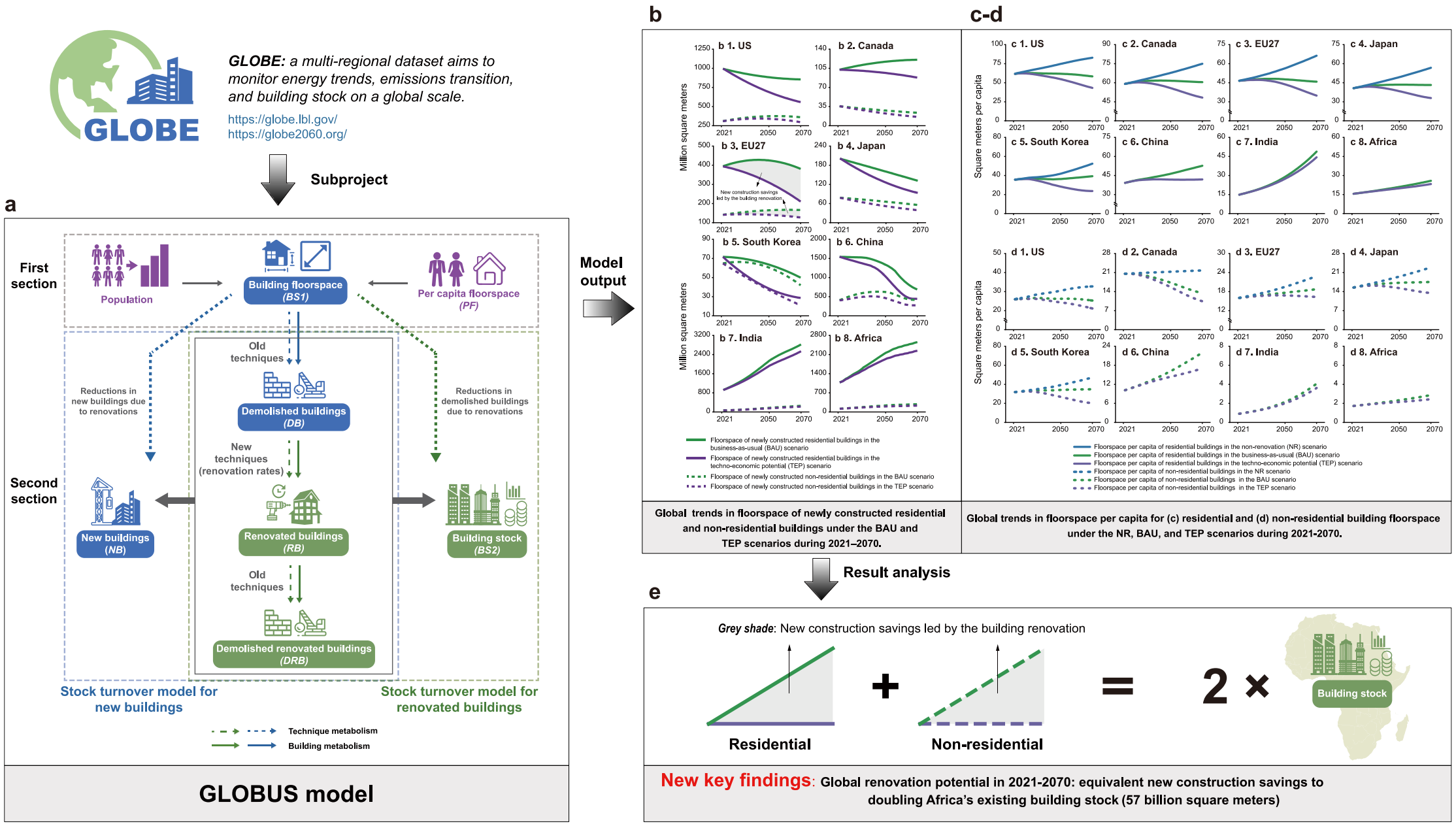

Potential impact of building renovation: A forecast suggests that global new construction could decrease by 57 billion square meters during 2021-2070, as evaluated by our global building stock model (GLOBUS), which integrates turnover and renovation dynamics.

## HIGHLIGHTS

- We created a global building stock model (GLOBUS) incorporating turnover and renovations.
- For many model scenarios, global building stock in 2070 will be 1.5-1.9 times that of 2020.
- For the BAU scenario, in 2070 building stocks in India and Africa will be 4 times those of 2020.
- China's building stock will increase at a compound average rate of 3 ‰/yr in 2020-2070.
- With each 1 ‰/yr renovation rate increase, global new construction falls by 123 million $m^2$/yr.

## SUMMARY

Buildings produce one-third of carbon emissions globally, however, data absence regarding global floorspace poses challenges in advancing building carbon neutrality. We compile the measured building stocks for 14 major economies and apply our global building stock model, GLOBUS, to evaluate future trends in stock turnover. Based on a scenario not considering renovation, by 2070 the building stock in developed economies will be ~1.4 times that of 2020 (100 billion $m^2$); in developing economies it is expected to be 2.2 times that of 2020 (313 billion $m^2$). Based on a techno-economic potential scenario, however, stocks in developed economies will decline to approximately 0.8 times the 2020 level, while stocks in developing economies will increase to nearly twice the 2020 level due to their fewer buildings currently. Overall, GLOBUS provides a way of calculating the global building stock, helping scientists, engineers, and policymakers conduct a range of investigation across various future scenarios.

## BROADER CONTEXT

Surpassing the two large emission sectors of transportation and industry, the building sector accounted for 34% and 37% of global energy consumption and carbon emissions in 2021, respectively. The building sector, the final piece to be addressed in the transition to net-zero carbon emissions, requires a comprehensive, multisectoral strategy for reducing emissions. Until now, the absence of data on global building floorspace has impeded the measurement of building carbon intensity (carbon emissions per floorspace) and the identification of ways to achieve carbon neutrality for buildings. For this study, we develop a global building stock model (GLOBUS) to fill that data gap. Our study's primary contribution lies in providing a dataset of global building stock turnover using scenarios that incorporate various levels of building renovation. By unifying the evaluation indicators, the dataset empowers building science researchers to perform comparative analyses based on floorspace. Specifically, the building stock dataset establishes a reference for measuring carbon emission intensity and decarbonization intensity of buildings within different countries. Further, we emphasize the sufficiency of existing buildings by incorporating building renovation into the model. Renovation can minimize the need to expand the building stock, thereby bolstering decarbonization of the building sector.

## INTRODUCTION

Global energy consumption and carbon emissions of the building sector exceeded even those of the major emission sectors of transportation[1] and industry[2], accounting for 34% and 37% of global energy consumption and carbon emissions in 2021, respectively.[3,4] Throughout a building's life cycle, emission levels are determined largely by building stock.[5,6] The potential doubling of the global building stock by mid-century would increase both end-use energy demand and operational carbon emissions by more than 50%.[7] Hence, strategically planning the scale of new constructions[8] and optimizing the use of existing buildings[9] are crucial to reducing building carbon emissions globally, and ultimately beginning a low-carbon transition in the building sector.[10]

A few previous studies have predicted a single country's building stocks for forecasting building carbon emissions. For example, the building stock in the United States (US) in 2100 is projected to be approximately 202% of the 2020 level.[11] Under a business-as-usual (BAU) scenario, China's per capita floorspace of non-residential buildings is expected to increase by more than 90% and reach 19.3 square meters ($m^2$)[*] by 2060.[12] India's non-residential building stock is projected to increase, from 1.1 billion $m^2$ in 2017 to 1.78 billion $m^2$ in 2027.[13] Moreover, while the prevailing perspective emphasizes the importance of refurbishing buildings with low-carbon materials and improving energy efficiency for building decarbonization, another impact of large-scale renovations is often overlooked: the avoidance of substantial carbon emissions from surplus new construction by increasing building lifetimes.[14]

### Our global building stock model

To bridge the research gaps in current studies, we develop a global building stock model (GLOBUS), a long-term forecasting model integrated with a turnover analysis that accounts for building renovations. The GLOBUS model is designed to simulate future changes in building stocks worldwide. We apply GLOBUS to address the following issues.

- What have been the global and regional trends in building floorspace since the millennium?
- How does the amount of newly built floorspace change when incorporating renovations?

---

[*] 1 square meter equals to 10.7639 square feet.

• How will global and regional building stocks change given different rates of renovation?

The GLOBUS model examines three scenarios: the non-renovation (NR), the BAU, and the techno-economic potential (TEP) scenarios. (These three scenarios are detailed in Section 3, Scenario Setting and Assumptions, of the Supplemental Information.) Initially, we estimate building stock based on the future population size and level of per capita floorspace in the NR scenario. This scenario incorporates the conventional life-cycle stages of building construction, operation, and demolition, excluding renovation. Results for the NR scenario answer the first question posed above, as calculated by Eq. 1 of the Experimental Procedures. When considering building renovation, two additional scenarios are incorporated: the BAU and TEP scenarios, which are based on variations in renovation rate*. The results of the turnover model that includes building renovations answer the second and third questions above. The detailed calculations and data sources for the BAU and TEP scenarios are found in the Experimental Procedures of this paper and Sections 2 & 3 of the Supplemental Information.

This study's greatest contribution is to establish a global building floorspace database for comparative analyses in building science research. Notably, the database provides the foundation for evaluating carbon intensity and decarbonization intensity within the building sector, quantified as carbon emissions per unit floorspace, aiding global comparisons of decarbonization levels among countries. Moreover, it highlights the practical importance of increasing building lifetime through renovation, which reduces both embodied and operational carbon emissions from surplus new building stock. Renovation preserves habitable floorspace, curbing the demand for new construction and demolition, thus ensuring that existing stock meets the needs of economic development.

* The BAU scenario posits a gradual annual increase in building renovations at the low rate countries currently are planning. The TEP scenario represents a swift and relatively ambitious growth in building renovations, representing a challenging yet aspirational trajectory.

## RESULTS

### Trends in global floorspace ignoring building renovation

Figure 1 illustrates the trends in building floorspace in 14 economies from 2000 to 2070 as estimated by the GLOBUS model using the NR scenario. (For trends in global floorspace per capita using the NR scenario, see Figure S1 of the Supplemental Information.) Economies that have large stocks of buildings are the US, the 27 countries of the European Union (EU27), China, India, Africa, Latin America and the Caribbean (LAC), and Indonesia, whose building floorspace is projected to reach tens of billions of square meters by 2070 (Figure 1 p). Because developed economies already possess substantial floorspace, projected building floorspace increases more slowly in developed economies (Figure 1 a-h) than in developing economies (Figure 1 i-n).

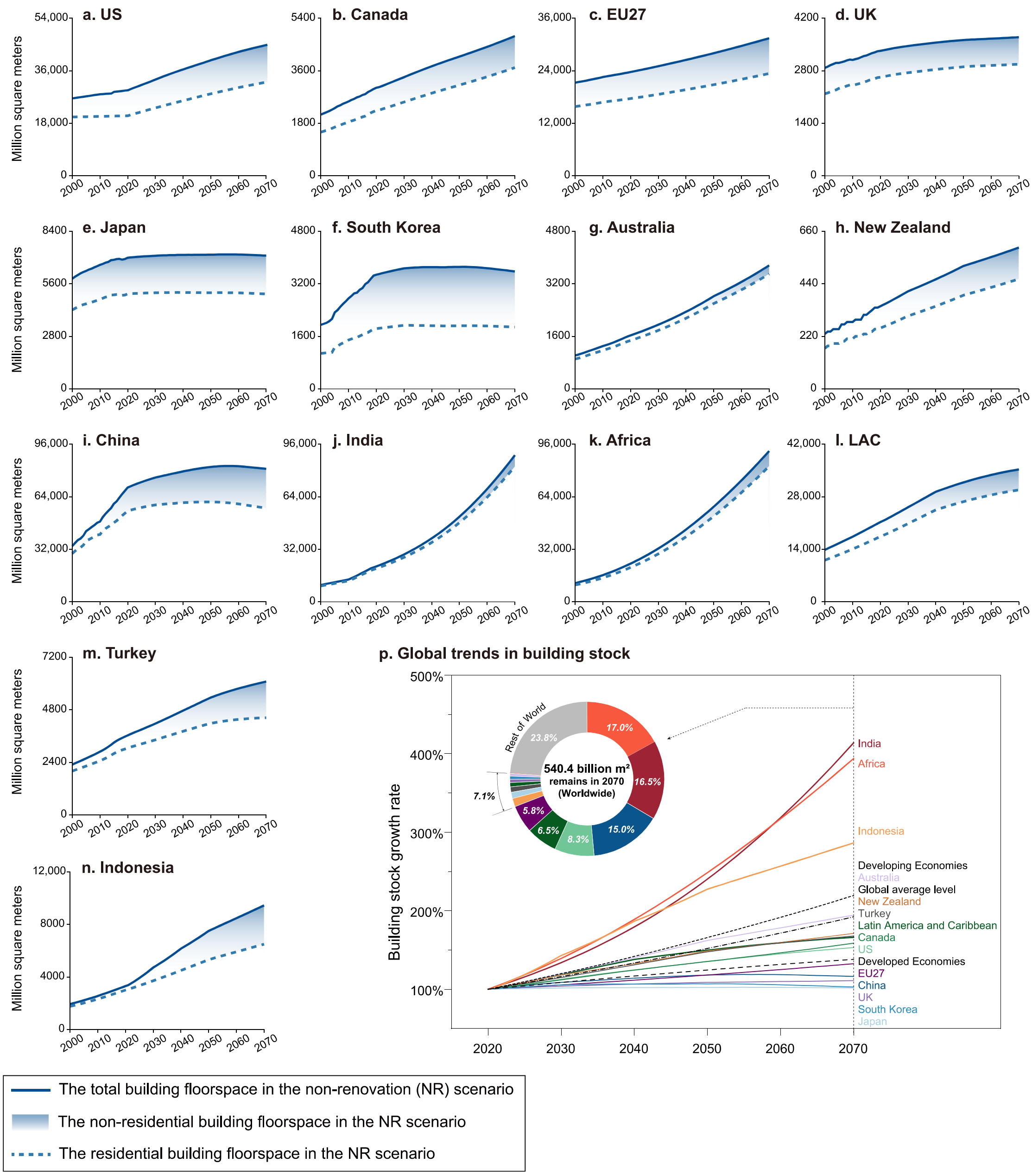

**Figure 1.** (a-n) Global trends in floorspace for residential and non-residential buildings; (p) global trends in building stock from 2000 to 2070 based on the NR scenario.

Based on the NR scenario, building floorspace in the US and EU27, here considered typical representatives of developed economies, will experience compound average annual growth rates of 0.8% and 0.6%, respectively, from 2000 to 2070. By 2070 total building floorspace in the US and EU27 will reach approximately 44.7 and 31.4 billion $m^2$, respectively, with residential buildings accounting for 71.7% and 74.4%, respectively, of that floorspace. Because of their low birth rates and aging populations, Japan and South

Korea will see future demands for building floorspace gradually slow, then decline and remain flat at approximately 7.1 and 3.6 billion m$^2$, respectively, after 2060. With respect to emerging economies, India and Africa will exhibit notable and swift expansion in building floorspace because of rapid population growth. Their total building floorspace will increase at a compound annual rate of 3.0% during 2000-2070, reaching approximately four times the floorspace in 2020. By 2070 the total building floorspace of India and Africa will reach 89.4 and 91.9 billion m$^2$, respectively, with residential buildings comprising 92.3% and 90.0%, respectively, of the total. Given China's declining population since the 2020s, future demand for building floorspace will decline: total building floorspace will increase by just 0.3% per year after 2020, arriving at 81.0 billion m$^2$ by 2070.

## Trends in global floorspace of future new construction

The assessment in Figure 1 did not consider the current status of building renovation in developed countries or the potential for renovation in developing countries. We used the GLOBUS model of building stock turnover to estimate the future floorspace of newly constructed buildings (see Figure 2) and future building stocks when accounting for various building renovation rates (see Figure 3).

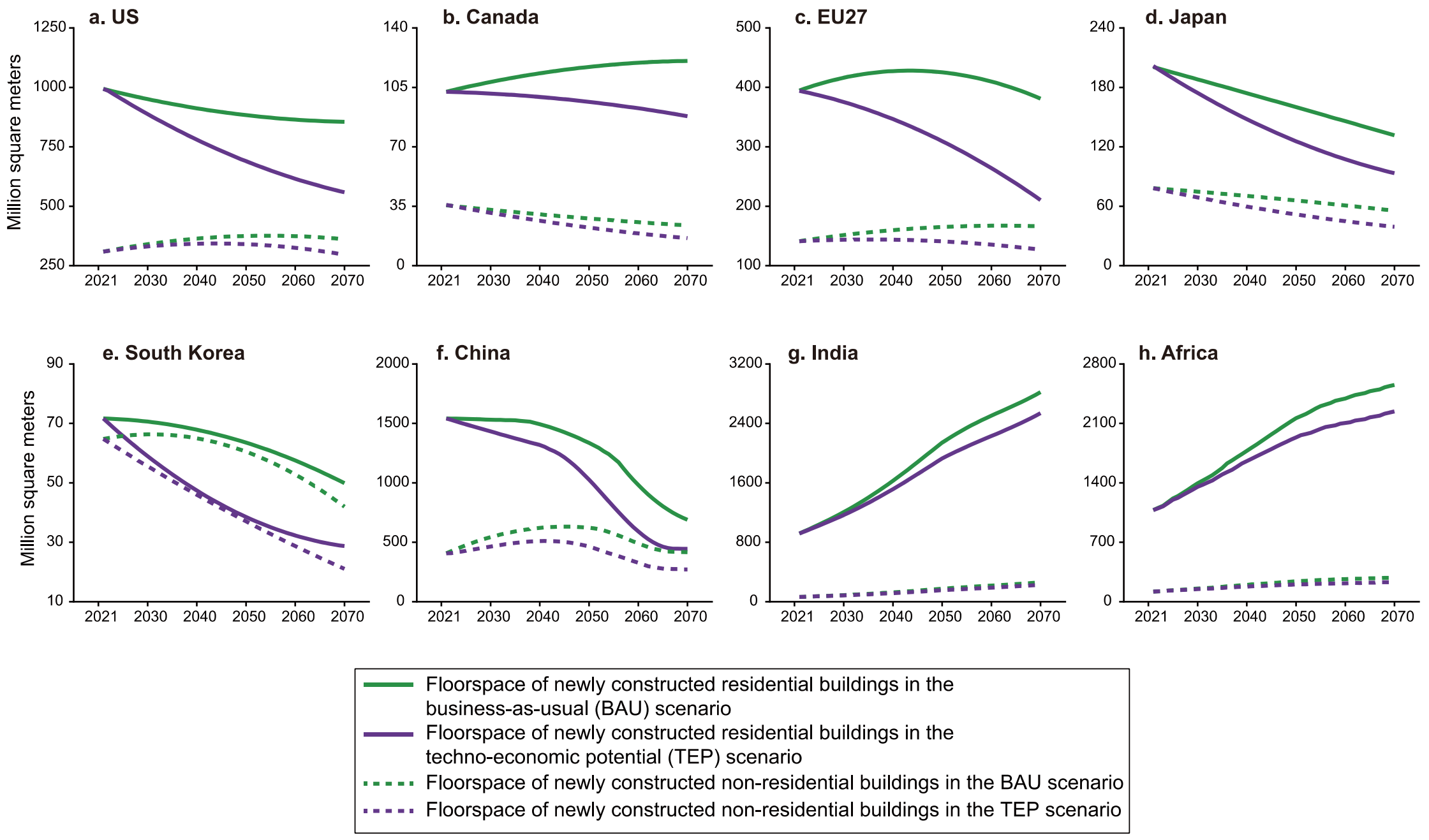

**Figure 2.** Trends in floorspace of newly constructed residential and non-residential buildings under the BAU and TEP scenarios in eight economies, 2021–2070.

As shown in Figure 2, even after incorporating building renovations, the US is still projected to have the greatest floorspace of new buildings in the eight economies. By 2070, newly constructed residential buildings in the US are estimated to reach 839.8 and 558.7 million $m^2$ for the BAU and TEP scenarios, respectively. By 2070 the floorspace of non-residential buildings in the US is expected to reach 361.3 and 279.7 million $m^2$ given the BAU and TEP scenarios, respectively. By 2070 Canada, Japan, and South Korea will be building no more than one-fifth as many new buildings as the US. By 2070 the floorspace of newly constructed buildings in the EU27 will rank second to those in the US. EU27 floorspace for residential buildings will reach 362.5 and 213.0 million $m^2$ based on the BAU and TEP scenarios, respectively, and for non-residential buildings will reach 166.0 and 129.3 million $m^2$. Meanwhile, by 2070, newly constructed residential buildings in India and Africa will represent more than triple the floorspace of those in the US given the BAU scenario (2821.5 and 2553.4 million $m^2$ in India and Africa, respectively), and be four times greater based on the TEP scenario in which the US will vigorously pursue building renovations. Based on the BAU scenario, by 2070 newly constructed floorspace of non-residential buildings is only about half that in the US (258.7 and 283.6 million $m^2$ in India and Africa, respectively). Given the TEP scenario floorspace approaches that in the US (226.7 and 232.9 million $m^2$ in India and Africa, respectively). Building renovation lowers newly constructed floorspace significantly in China: the new construction is projected to reach approximately 684.0 and 432.4 million $m^2$ in residential buildings and 415.4 and 271.7 million $m^2$ in non-residential buildings under the BAU and TEP scenarios, respectively, by 2070.

**Potential changes in per capita floorspace given building renovation**

Figure 3 shows the per capita floorspaces of the eight selected countries under the NR, BAU, and TEP scenarios given various building renovation rates. (Regarding future building stock changes that account for building renovation, see Figure S2 of the Supplemental Information.)

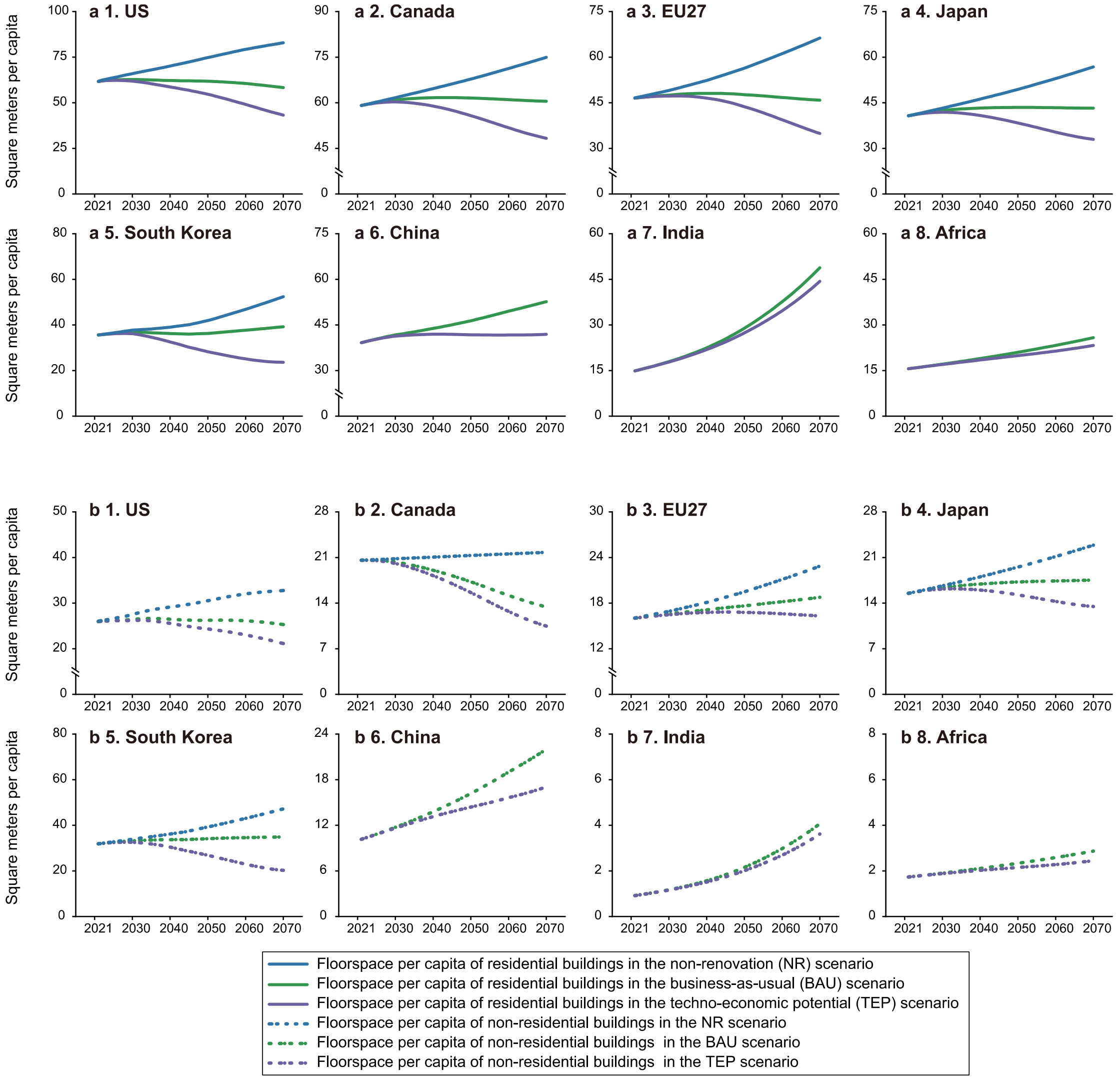

**Figure 3.** Trends in floorspace per capita for (a) residential and (b) non-residential buildings for eight economies under the NR, BAU, and TEP scenarios, 2021-2070.

The per capita floorspace of residential buildings in the US and Canada is much greater than that in other economies (see Figure 3 a). By 2070, per capita floorspace in the US will be 58.3 and 43.2 $m^2$ for the BAU and TEP scenarios, respectively, while that in Canada will reach 60.5 and 48.3 $m^2$ based on the BAU and TEP scenarios, respectively. By 2070 per capita floorspace given the BAU and TEP scenarios will decrease by 29.6% and 47.8%, respectively, and in Canada will decrease by 19.3% and 35.5%, respectively, compared with results under the NR scenario. In the EU27, which has an intermediate level of per capita floorspace, by 2070 per capita floorspace in residential buildings will decline by more than 20 and 30 $m^2$ under the BAU and TEP scenarios, respectively, compared with the NR scenario. By 2070 the per capita floorspace of residential buildings in the EU27 will reach 45.8 and 34.9 $m^2$ under the BAU and TEP scenarios, respectively. By 2070 per capita floorspace in Japan and South Korea will reach 43.2 and 39.2 $m^2$, respectively, under the BAU scenario and 33.0 and 23.6 $m^2$, respectively, under the TEP scenario; levels based on the NR scenario are 56.8 and 52.4 $m^2$, respectively. As for developing economies, the per capita floorspace of residential buildings in China will decrease to approximately 41.9 $m^2$ by 2070 given the TEP scenario compared with 52.6 $m^2$ for the NR scenario. Although under the BAU and TEP scenarios per capita floorspace in most developed economies maintains a stable or even declining trend compared with 2021 values, in India and Africa per capita floorspace maintains relatively rapid growth under both scenarios. Based on the TEP scenario, for example, per capita floorspace of India and Africa will reach 44.4 and 23.3 $m^2$, respectively, by 2070.

According to the results for non-residential buildings shown in Figure 3 b, per capita floorspace in 2070 in South Korea will reach 34.9 $m^2$ under the BAU scenario and 20.2 $m^2$ for the TEP scenario. Those are the highest levels of per capita floorspace in the eight countries for countries that show the greatest declines compared to the NR scenario (decreases of 12.3 and 27.0 $m^2$ in per capita floorspace for the BAU and TEP scenarios, respectively). Among the eight economies, the US has a relatively spacious per capita floorspace of 25.3 and 21.1 $m^2$ based on the BAU and TEP scenarios, respectively, by 2070. The per capita floorspace for the EU27 and Japan in 2070 are similar: approximately 18.8 and 17.5 $m^2$ under the BAU scenario and 16.3 and 13.5 $m^2$ under the TEP scenario, respectively. In addition, by 2070 Canada will have approximately 13.4 $m^2$ per capita of non-residential building floorspace for the BAU scenario and 10.5 $m^2$ per capita given the TEP scenario. Among developing countries, the per capita floorspace of non-residential buildings in China is the highest, at 17.0 $m^2$ under the TEP scenario. The per capita non-residential floorspace for India and Africa is only 3.6 and 2.5 $m^2$, respectively, based on the TEP scenario. As with residential buildings, as the rate of renovation increases, the per capita floorspace in developed countries stabilizes under the BAU scenario or decreases under the TEP scenario compared to the 2021 level. The per capita floorspace in developing countries, however, continues to increase to varying degrees.

## DISCUSSION

We focused on several implications of the GLOBUS model results from the three scenarios. Topics are tracking the carbon intensity of building operations and achieving sustainable development of the future building stock.

### Carbon intensity of global building operations

Floorspace plays a crucial role in measuring carbon emission trends and decarbonization levels of buildings, serving as a key indicator of carbon intensity in terms of carbon emissions per floorspace. For instance, Figure 4 a-i illustrates the variations in total carbon emissions, carbon emissions per capita, and carbon emissions per floorspace released by residential building operations in 11 economies in 2000, 2011, and 2021. Generally, levels of carbon emissions per floorspace and per capita for residential buildings tend to be greater in developed countries than in developing countries. For example, in 2021 the carbon emissions per floorspace in the US (45.2 kilograms of carbon dioxide per square meter [$kgCO_2/m^2$]) were three times greater than those in China (14.5 $kgCO_2/m^2$) and more than twice those in India (18.5 $kgCO_2/m^2$). The carbon emissions per capita in the US (2797.3 kilograms of carbon dioxide per capita [$kgCO_2$/person]) were five times greater than those in China (566.6 $kgCO_2$/person) and more than ten times greater than those in India (275.7 $kgCO_2$/person).

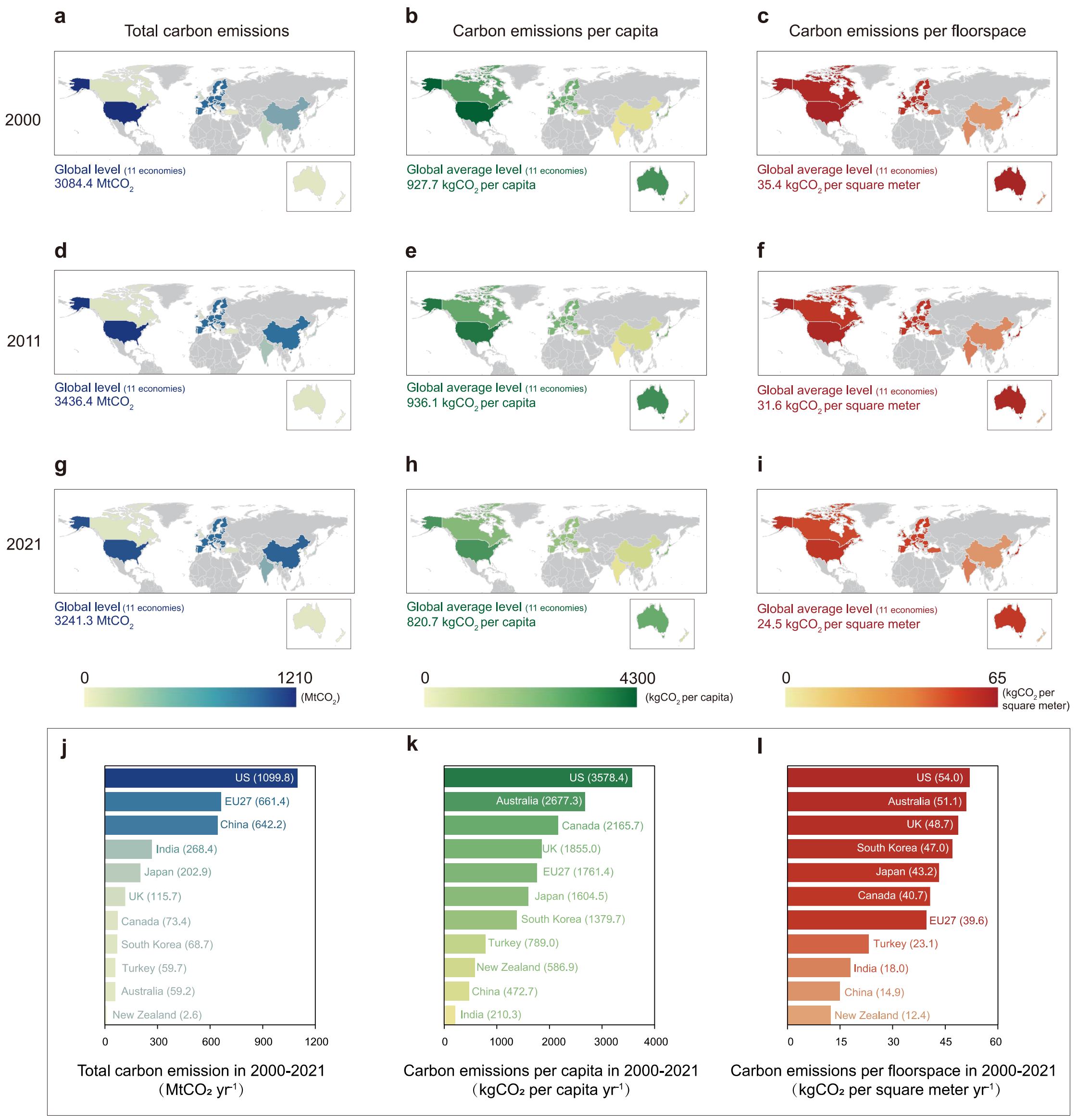

**Figure 4.** Carbon emissions from residential building operations in 11 economies from 2000 to 2021. Note: the carbon emission data were referred from our former study[15] and our GLOBE database (https://globe.lbl.gov/, https://globe2060.org/).

Figure 4 j-l reveals two facets of the annual average results between 2000 and 2021: On one hand, the US exhibited the highest total, per capita, and per floorspace carbon emissions during that period; on the other hand, although China and India ranked among the top four in total carbon emissions, their per capita and per floorspace emissions were among the lowest three. The above observations underscore the opportunities and challenges facing every economy, especially every emerging economy: The large labor force that a sizable population provides is conducive to industrial growth, but also poses difficulties, particularly as regards mitigating energy-related carbon emissions. In another word, it is challengeable for emerging economies to balance energy decarbonization and economic growth in the upcoming decades. As for non-residential building operations in the 11 economies from 2000 to 2021, their total carbon emissions, carbon emissions per capita, and carbon emissions per floorspace are illustrated in Figure S3 of the Supplemental Information.

## Sustainable development of future building stocks

Buildings have long life cycles, require large investments,[16] and during construction rely heavily on high-emission manufacturers of steel, cement, and other building materials, creating substantial carbon lock-in.[17] Buildings then depend on the electricity industry during operation, making it difficult to lower emissions significantly by only implementing energy-saving measures.[18,19] Thus we emphasize the sufficiency of the existing building stock. Because the planning stage of construction may either conserve resources or produce a cascade of consequences that are not cost-effective or sustainable, planning should maximize the utilization of existing buildings and minimize the construction of new buildings. In all countries the thoughtful, sustainable development of building stocks deserves increased attention in the effort to achieve the deep decarbonization of buildings.

We recommend that, to achieve sustainable development of building stocks, governments consider the following three factors when formulating policies and initiating associated measures. First, the scale of newly constructed buildings should be planned from the demand side to avoid unfettered reconstruction while ensuring adequate floorspace for economic development. Second, governments should apply measures such as renovation appropriately to improve the energy efficiency of existing buildings rather than automatically carrying out large-scale demolition and reconstruction. Third, recyclable low-carbon emission building materials should be chosen, clean construction practices should be promoted, and prefabricated buildings should be encouraged to decrease on-site construction time, construction waste, and labor costs. Additionally, our study delved into the implications of the GLOBUS model results, particularly in the context of mitigating embodied carbon through building renovation. We specifically examined the case of China's building sector and the pertinent discussion is provided in Figure S4 of the Supplemental Information.

## Conclusions

Using the GLOBUS model, we forecasted trends in building stocks based on three scenarios that incorporate different rates of building renovation. Our analysis involved 14 major economies during the period 2000 to 2070. For the NR scenario, the global building floorspace is projected to increase steadily. Results for the BAU and TEP scenarios differ: by 2070 newly constructed residential floorspace in, for instance, India or Africa will be more than triple that of the US based on the BAU scenario and four times that given the TEP scenario. The newly constructed non-residential floorspace of India or Africa in 2070 will be approximately half of that in the US under the BAU scenario and close to the US level under the TEP scenario.

Incorporating building renovation maximizes the utilization of existing building stock and lessens the expansion of new construction. Compared with 2021 most developed economies maintain a stable or even declining trend in per capita floorspace of residential buildings based on both the BAU and TEP scenarios. Per capita floorspace in India and Africa, on the other hand, demonstrates rapid growth under both scenarios. The per capita floorspace of non-residential buildings in developing countries also will continue to increase, even given a BAU or TEP scenario.

We posit that estimating the global building stock facilitates comparisons of building carbon emissions based on levels of carbon intensity. Additionally, providing thoughtful planning for new building construction and improving the sufficiency of existing buildings can foster the sustainable development of building stocks. We emphasize that existing buildings, when renovated, can minimize the need to expand the building stock and support decarbonization of the building sector.

In summary, our study's primary contribution lies in creating an initial dataset of global building stocks that can be used to assess stock turnover based on scenarios that involve various rates of building renovation. Using our dataset and model, scientists, engineers, and policymakers can perform comparative analyses per floorspace. By harmonizing the evaluation parameters for comparative analysis, our GLOBUS establishes a reference for measuring the changes in carbon intensity of buildings (quantified by carbon emissions per floorspace) and corresponding decarbonization intensity among countries.

## EXPERIMENTAL PROCEDURES

### Overview of our GLOBUS model

We developed the GLOBUS model to examine global building stocks from 2000 to 2070. The GLOBUS consists of two major sections (see Figure S5 in the Supplemental Information): a building stock analysis under the NR scenario and an analysis that considers various building renovation rates. The analysis performed by the second section, which is an extension of that performed by the first, incorporates analysis of the BAU and TEP scenarios.

### Stock analysis for the non-renovation scenario

The first section of the GLOBUS analyzes building stocks based on the NR scenario, starting with the prediction of per capita floorspace ($\boldsymbol{PF}$) as its foundational element. Drawing from historical data on global building stocks, we derived the potential building stock ($\boldsymbol{BS}$) by forecasting each economy's population and per capita floorspace:

$$\boldsymbol{BS^{NR}_{i,j,t}} = \boldsymbol{PF_{i,j,t}} \times \boldsymbol{Population_{i,j,t}} \quad \text{(Eq. 1)}$$

where $\boldsymbol{i}$ represents building types (residential and non-residential); $\boldsymbol{j}$ expresses various economies; and $\boldsymbol{t}$ indicates the evaluation period of 2000–2070.

The population data were obtained from https://www.populationpyramid.net/ and Eurostat.[20] When predicting per capita floorspace, we sought validation from reputable and widely acknowledged institutions; from databases (International Energy Agency,[21-23] U.S. Energy Information Administration,[24] United Nation Environment Programme,[25] and Eurostat [20]); or high-quality peer-reviewed papers authored by experts native to relevant countries.[26-28] Detailed reference information regarding per capita floorspace in each country is provided in Section 3, Scenario Setting and Assumptions, of the Supplemental Information.

### Stock analysis including turnover

Although analysis by the NR scenario developed initial potential stocks based on per capita floorspace, it did not account for renovated buildings ($\boldsymbol{RB}$). Renovation can increase the lifetimes,[29] while mitigating the demand for new buildings ($\boldsymbol{NB}$). Considering building renovation, the calculation of the potential floorspace of new buildings is outlined as follows.

$$\boldsymbol{NB_{i,j,t}^{BAU}} = \boldsymbol{BS_{i,j,t}^{NR}} - \boldsymbol{BS_{i,j,t-1}^{NR}} + \boldsymbol{DB_{i,j,t}^{BAU}} - \boldsymbol{RB_{i,j,t}^{BAU}} + \boldsymbol{DRB_{i,j,t}^{BAU}} \quad \text{(Eq. 2)}$$

$$\boldsymbol{NB_{i,j,t}^{TEP}} = \boldsymbol{BS_{i,j,t}^{NR}} - \boldsymbol{BS_{i,j,t-1}^{NR}} + \boldsymbol{DB_{i,j,t}^{TEP}} - \boldsymbol{RB_{i,j,t}^{TEP}} + \boldsymbol{DRB_{i,j,t}^{TEP}} \quad \text{(Eq. 3)}$$

where $\boldsymbol{DB}$ denotes demolished buildings, signifying the routine demolition of a building when it reaches the end of its service lifespan, and $\boldsymbol{DRB}$ indicates renovated buildings that are demolished. For detailed calculation instructions (Eqs. S1-S5) and scenario assumptions, refer to Sections 2 & 3 of the Supplemental Information, respectively.

Although the GLOBUS model assessed the floorspace of new buildings for the BAU and TEP scenarios given various renovation rates (see Tables S1 and S2 of the Supplemental Information), the assessment is based on maintaining the building stock calculated in the NR scenario (i.e., $\boldsymbol{BS_{i,j,t}^{NR}}$). When building renovation extends the lifespan of a building to satisfy some of the building demand, the overall demand is reduced. [30] The reduction should be reflected directly in the building stock, and can be expressed quantitatively as a reduction in the building stock estimated by the NR scenario ($\boldsymbol{BS_{i,j,t}^{NR}}$). The representation of building stocks at various renovation levels is as follows.

$$\boldsymbol{BS_{i,j,t}^{BAU}} = \boldsymbol{BS_{i,j,t}^{NR}} - \sum(\boldsymbol{RB_{i,j,t}^{BAU}} - \boldsymbol{DRB_{i,j,t}^{BAU}}) \quad \text{(Eq. 4)}$$

$$\boldsymbol{BS_{i,j,t}^{TEP}} = \boldsymbol{BS_{i,j,t}^{NR}} - \sum(\boldsymbol{RB_{i,j,t}^{TEP}} - \boldsymbol{DRB_{i,j,t}^{TEP}}) \quad \text{(Eq. 5)}$$

## SUPPLEMENTAL INFORMATION

The supplemental materials appear in the e-component of this submission.

## ACKNOWLEDGMENTS

This manuscript has been authored by an author at Lawrence Berkeley National Laboratory under Contract No. DE-AC02-05CH11231 with the U.S. Department of Energy. The U.S. Government retains, and the publisher, by accepting the article for publication, acknowledges, that the U.S. Government retains a non-exclusive, paid-up, irrevocable, world-wide license to publish or reproduce the published form of this manuscript, or allow others to do so, for U.S. Government purposes.

## AUTHOR CONTRIBUTIONS

Conceptualization, M.M., N.Z., and J.Y.; Methodology, M.M., S.Z., R.Y., and K.Y.; Software, S.Z., R.Y., and K.Y.; Validation, J.Z., and J.K.; Writing-Original Draft, M.M., and S.Z.; Writing-Review & Editing, M.M., N.Z., J.Y., and W.F.; Funding Acquisition, N.Z.; Supervision, M.M., N.Z., and J.Y..

## DECLARRATION OF INTERESTS

The authors declare no competing interests.